\documentclass[aps,amsmath,amssymb,twocolumn]{revtex4-1}

\usepackage{graphicx}
\usepackage{dcolumn}
\usepackage{bm}
\usepackage{color}
\usepackage[normalem]{ulem}

\begin{document}

\title{Near-field radiative heat transfer between one-dimensional magneto-photonic crystals}

\author{E. Moncada-Villa$^{1}$}
\author{J.~C. Cuevas$^{2}$}

\affiliation{$^{1}$Escuela de F\'{\i}sica, Universidad Pedag\'ogica y Tecnol\'ogica de Colombia,
Avenida Central del Norte 39-115, Tunja, Colombia}

\affiliation{$^2$Departamento de F\'{\i}sica Te\'orica de la Materia Condensada
and Condensed Matter Physics Center (IFIMAC), Universidad Aut\'onoma de Madrid,
E-28049 Madrid, Spain}

\date{\today}

\begin{abstract}
We present a theoretical study of the effect of an external dc magnetic field in the near-field radiative heat transfer between two 
one-dimensional magneto-photonic crystals with unit cells comprising a magneto-optical layer made of $n$-doped InSb and a dielectric 
layer. We find that in absence of an external field, and depending on the gap size, the radiative heat transfer between these
multilayer structures can be larger or smaller than that of the case of two InSb infinite plates. On the other hand, when an external 
magnetic field is applied, the near-field radiative heat transfer is reduced as a consequence of the suppression of hybridized surface 
polariton waves that are supported for transverse magnetic polarized light. We show that such reduction is exclusively due to the 
appearance of magnetic-field induced hyperbolic modes, and not to the polarization conversion in this magneto-optical system.
\end{abstract}

\maketitle

\section{Introduction} \label{sec-intro}

In recent years we have witnessed a true revolution in the field of thermal radiation \cite{Song2015a,Cuevas2018,Biehs2020}. 
This has been mainly triggered off by experimental advances that, in particular, have made possible to confirm the long-standing 
prediction that the limit set by Stefan-Boltzmann's law for the radiative heat transfer between two bodies can be largely overcome 
by bringing them sufficiently close \cite{Polder1971}. In the near-field regime, i.e., when the separation between two bodies 
is smaller than the thermal wavelength $\lambda_{\rm Th}$ ($\sim$10 $\mu$m at room temperature), they can also exchange radiative 
heat via evanescent waves (or photon tunneling), which are not taken into account in the derivation of Stefan-Boltzmann's law. 
This additional contribution can completely dominate the near-field radiative heat transfer (NFRHT) and, in turn, can
lead to overcome the Stefan-Boltzmann's limit (or blackbody limit) by orders of magnitude for sufficiently small gaps. There is
by now an overwhelming experimental evidence of this fact that has obtained in a great variety of systems and using many different materials 
\cite{Kittel2005,Narayanaswamy2008,Hu2008,Rousseau2009,Shen2009,Shen2012,Ottens2011,Kralik2012,Zwol2012a,Zwol2012b,Guha2012,
Worbes2013,Shi2013,St-Gelais2014,Song2015b,Kim2015,Lim2015,St-Gelais2016,Song2016,Bernardi2016,Cui2017,Kloppstech2017,Ghashami2018,
Fiorino2018,DeSutter2019}. From a fundamental point of view, these experiments have also helped to firmly establish the theory
of fluctuational electrodynamics \cite{Rytov1953,Rytov1989}, which has become the standard model for the description of NFRHT. 
From a more applied point of view, NFRHT may have an important impact in different technologies that make use of thermal radiation,
see Refs.~\cite{Song2015a,Cuevas2018,Biehs2020} and references therein.

The basic physical mechanisms underlying NFRHT are relatively well-understood by now. In this sense, a good part of the efforts of the
theory community in the field of thermal radiation focuses now on finding strategies to actively control NFRHT with the hope to
develop novel thermal functional devices. Many interesting ideas have been put forward in recent years, and the interested reader
can consult the recent reviews of Refs.~\cite{Song2015a,Cuevas2018,Biehs2020}. One of the most attractive and promising ideas is the
use an external magnetic field to control the NFRHT between magneto-optical (MO) materials, which we proposed some years ago in
Ref.~\cite{Moncada-Villa2015}. In that work, we showed that the NFRHT between two parallel plates made of doped semiconductors can 
be substantially altered by the application of a static magnetic field. This work paved the way for the prediction of a plethora of 
thermomagnetic effects. Thus, for instance, it has been predicted that the lack of reciprocity in MO systems can lead to novel 
phenomena such as a near-field thermal Hall effect \cite{Ben-Abdallah2016a} or the existence of a persistent heat current \cite{Zhu2016}. 
It has also been theoretically demonstrated that MO systems under a static magnetic field can exhibit many phenomena that are 
the near-field thermal analogues of basic transport effects in the field of spintronics such as giant thermal magnetoresistance 
\cite{Latella2017} or anisotropic thermal magnetoresistance \cite{Abraham-Ekeroth2018,Moncada-Villa2020}. Many of these phenomena 
have been reviewed in Refs.~\cite{Ott2019,Biehs2020}.

Most theoretical studies of the effect of a magnetic field in the NFRHT between MO systems have shown that the field tend
to reduce the NFRHT and only a modest enhancement was found in Ref.~\cite{Moncada-Villa2020} in a very asymmetric situation.
The reason for this reduction in extended systems (like infinite parallel plates) is that a magnetic field induces the 
appearance of hyperbolic modes that replace the surface modes, both surface plasmon polaritons (SPPs) and surface phonon
polaritions (SPhPs), that dominate the NFRHT in these systems in the absence of magnetic field \cite{Moncada-Villa2015,Moncada-Villa2019}. 
These hyperbolic modes, in spite of being propagating inside the material, are less effective than surface modes transferring
the heat across the gap between two bodies. In this regard, the recent prediction put forward in Ref.~\cite{Song2020} that an
external magnetic field can greatly enhance the NFRHT between two multilayer hyperbolic metamaterials (consisting of alternating layers
of a MO material and a dielectric) is certainly very interesting, albeit quite surprising. There has been quite some attention
devoted to this type of multilayer systems in recent years \cite{Guo2012,Biehs2012,Guo2013,Biehs2013,Bright2014,Miller2014,Biehs2017,
Iizuka2018}. The main reason for this interest is that the contribution of surface modes at multiple interfaces can greatly 
enhance the NFRHT, as compared to the case of two infinite parallel plates, see e.g.\ Ref.~\cite{Iizuka2018}. The conclusion that
an external magnetic can enhance the NFRHT between one-dimensional (1D) magneto-photonic crystals was reached in Ref.~\cite{Song2020}
making use of an effective medium theory, a type of theory that is known to have limitations in the description of NFRHT 
\cite{Biehs2017,Iizuka2018,Fernandez-Hurtado2017}. In this regard, it would be highly desirable to revisit this interesting problem 
making use of an exact approach. This is precisely the main goal of this work.

\begin{figure}[t]
\includegraphics[width=\columnwidth,clip]{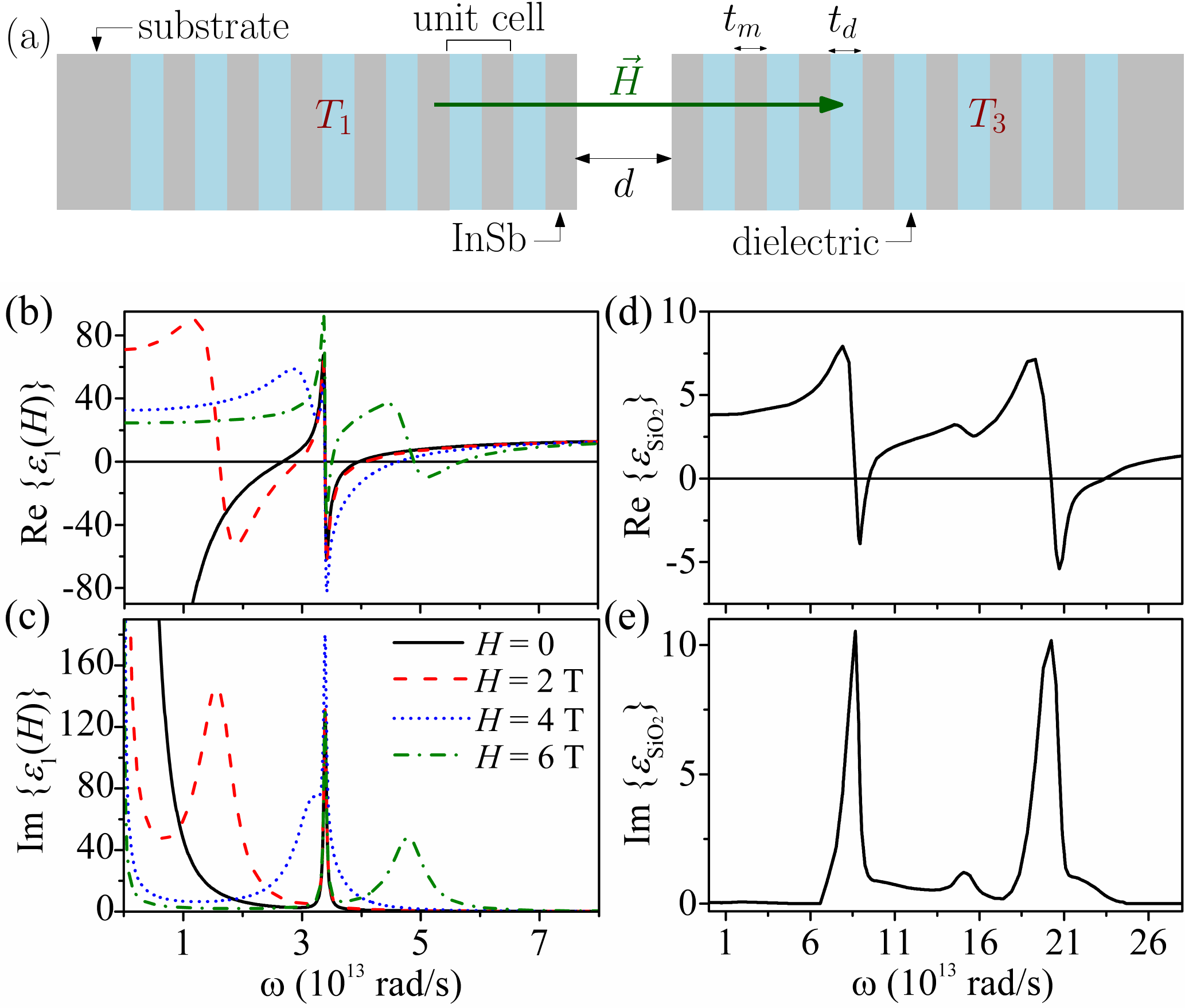}
\caption{(a) Schematic view of two identical periodic multilayer structures, separated by a vacuum gap of thickness $d$, each one 
at its corresponding temperature $T_1$ and $T_3$, and under the influence of an external dc magnetic field $H$ oriented along the 
stratification ($z$) axis. The unit cells comprise a layer of InSb, with thickness $t_m$, and a dielectric layer such as BK7 or SiO$_2$ of 
thickness $t_d$ (assumed to be 10 nm in all calculations). Both structures are grown on InSb substrates. In panels (b) and (c), 
we show, the real and imaginary part, respectively, of the diagonal element of InSb dielectric tensor for different values of the external 
magnetic field. In panels (d) and (e) we present the real and imaginary part, respectively, of the dielectric function of SiO$_2$.}
\label{fig-system}
\end{figure}

In this paper we present a systematic theoretical study of the influence of an external magnetic field in the NFRHT between two identical
1D magneto-photonic crystals with unit cells comprising a MO layer, made of InSb, and a dielectric layer, made of either BK7 glass
or SiO$_2$, see Fig.~\ref{fig-system}(a). Our calculations are based on a scattering matrix approach for anisotropic materials that provides 
the exact numerical results for an arbitrary number of layers in these crystals. We find that, irrespective of the different geometrical 
parameters, such as gap size, thickness of the dielectric layer, etc., and irrespective the dielectric material (BK7 glass or silica), an 
external magnetic field always reduces the NFRHT as compared to the case of two InSb parallel plates. In particular, we find that for a
field of a few Teslas the NFRHT of these layered systems can be reduced by almost a factor of 6, as compared to the zero-field
case. Moreover, we show that the physical mechanism for this drastic magnetic-field reduction is the appearance of hyperbolic modes
that replace the hybrid surface polariton modes supported by these metal-dielectric multilayer structures.  
    
The remainder of this paper is organized as follows. In Sec.~\ref{sec-theory} we describe the 1D magneto-photonic crystals under 
study, as well as the theoretical approach used for the calculation of the radiative heat transfer between them. 
In Sec.~\ref{sec-zero}, we start the discussion of our main results by analyzing the radiative heat transfer in the absence of an 
applied field, paying special attention to the reduction of the heat flux in an intermediate regime of the vacuum gap size, and to 
the enhancement of radiative heat flux in the far-field regime. Then, in Sec.~\ref{sec-finite}, we analyze the effect of 
an external magnetic field in the NFRHT between these multilayer structures. Finally, in Sec.~\ref{sec-conclusions} we summarize 
the main results presented in this manuscript.

\section{System under study and theoretical approach} \label{sec-theory}

The goal of this work is the theoretical description of the radiative heat transfer between two identical multilayer structures, 
both deposited on InSb substrates, separated by a vacuum gap of size $d$ and under the action of an external dc magnetic field 
of magnitude $H$ [see Fig.~\ref{fig-system}(a)]. Each multilayer comprises $N$ unit cells with a dielectric layer of either SiO$2$ 
or BK7 glass, of thickness $t_d$, and a MO layer of InSb, with thickness $t_m$. The external magnetic field, assumed to 
be parallel to the stratification axis ($z$-axis), generates a MO activity inside the InSb material \cite{Zvezdin1997}, which is 
described by the following dielectric tensor \cite{Palik1976} 
\begin{equation}
\label{perm-tensor-theta}
\hat \epsilon = 
\left( \begin{array}{ccc} 
\epsilon_1  &  -i\epsilon_2  &  0 \\
i\epsilon_2 &   \epsilon_1   &  0 \\
           0      &0 &  \epsilon_3 
\end{array} \right) ,
\end{equation}
where
\begin{eqnarray}
\epsilon_1(H) & = & \epsilon_{\infty} \left( 1 + \frac{\omega^2_L - \omega^2_T}{\omega^2_T - 
\omega^2 - i \Gamma \omega} + \frac{\omega^2_p (\omega + i \gamma)}{\omega [\omega^2_c -
(\omega + i \gamma)^2]} \right) , \nonumber \\
\label{eq-epsilons}
\epsilon_2(H) & = & \frac{\epsilon_{\infty} \omega^2_p \omega_c}{\omega [(\omega + i \gamma)^2 -
\omega^2_c]} , \\
\epsilon_3 & = & \epsilon_{\infty} \left( 1 + \frac{\omega^2_L - \omega^2_T}{\omega^2_T -
\omega^2 - i \Gamma \omega} - \frac{\omega^2_p}{\omega (\omega + i \gamma)} \right) . \nonumber
\end{eqnarray}
Here, $\epsilon_{\infty}$ is the high-frequency dielectric constant, $\omega_L$ ($\omega_T$ ) is  the longitudinal 
(transverse) optical phonon frequency, $\omega^2_p =ne^2/(m^{\ast} \epsilon_0  \epsilon_{\infty})$ is the plasma frequency of free 
carriers of density $n$ and effective mass $m^{\ast}$, $\Gamma$ ($\gamma$) is the phonon (free-carrier) damping constant, and 
$\omega_c =  eH/m^{\ast}$ is the cyclotron frequency, which depends on the intensity of the external magnetic  field. All the 
calculations presented in this work were done assuming the following parameter values taken from Ref.~\cite{Palik1976}: 
$\epsilon_{\infty}  = 15.7$, $\omega_L = 3.62 \times 10^{13}$ rad/s, $\omega_T = 3.39\times 10^{13}$ rad/s,  
$\Gamma = 5.65 \times 10^{11}$ rad/s, $\gamma = 3.39 \times 10^{12}$ rad/s, $n = 1.07 \times 10^{17}$  cm$^{-3}$, 
$m^{\ast}/m = 0.022$, $\omega_p = 3.14 \times 10^{13}$ rad/. Moreover, we set the thickness of the dielectric layers to $t_d=10$ nm.
We present in Fig.~\ref{fig-system}(b,c) the real and imaginary parts of the component $\epsilon_1(H)$ for such a choice of 
parameters and different values of the magnitude of the applied field. According to Eq.~\eqref{eq-epsilons}, in the absence of 
an external field it follows that $\epsilon_1(H=0)=\epsilon_3$. However, as can be observed in Fig.~\ref{fig-system}(b), the 
application of an external field leads to the appearance of frequency regions in which $\epsilon_1(H)$ and $\epsilon_3$ have opposite 
signs. These regions are characterized by the existence of electromagnetic modes, known as hyperbolic modes, that are propagating 
inside the InSb and evanescent out of it. These modes are responsible of the progressive disappearance of the SPP modes occurring 
below the surface plasmon frequency $\omega_{sp}=\omega_p/\sqrt{2}$, and of the SPhP modes occurring between the longitudinal and 
transverse optical frequencies $\omega_T$ and $\omega_L$, as it has been shown for infinite parallel plates and thin films made of 
InSb \cite{Moncada-Villa2015,Moncada-Villa2019}.

In this work, we consider two possible materials for the dielectric layers. The first one is the BK7 glass, with a constant (and real) 
dielectric function of $\varepsilon_d=2.25$. The second one is SiO$_2$, a polar material whose dielectric function has the real and 
imaginary parts presented in panels (d) and (e) of Fig.~\ref{fig-system} that were taken from Ref.~\cite{Palik1985}. As it is well-known,
for frequencies in which the real part of this dielectric function is negative, an interface between an infinite plate of this material 
and vacuum can support SPhPs whose evanescent field contributes substantially to the NFRHT transfer \cite{Mulet2002,Shen2009,Song2015b}.  

As explained above, we focus here on the calculation of the radiative heat flux between two 1D magneto-photonic 
crystals [see Fig.~\ref{fig-system}(a)]. This calculation was performed within the framework of the fluctuational electrodynamics
theory \cite{Rytov1953,Rytov1989}. In particular, we are interested in the radiative linear heat conductance per unit of area or heat 
transfer coefficient, $h_N$, between the two anisotropic multilayer systems, each one at its corresponding temperature $T_1$ and 
$T_3$ [see Fig.~\ref{fig-system}(a)]. This heat transfer coefficient is defined in terms of the net power per unit of area exchanged 
between two anisotropic layered systems, $Q_N$, via the relation  
\begin{equation}
h_N(T,d) = \lim_{\Delta T \rightarrow 0^+} \frac{Q_N(T_1=T+\Delta T,T_3=T,d)}{\Delta T} , 
\end{equation}
where \cite{Biehs2011}
\begin{equation}
\label{eq-net-Q}
Q_N = \int^{\infty}_{0} \frac{d \omega}{2\pi} \left[ \Theta_1(\omega) - \Theta_3(\omega) 
\right] \int \frac{d{\bf k}}{(2\pi)^2} \tau(\omega,{\bf k},d). 
\end{equation}
In this Landauer-like expression, $\Theta_i(\omega) = \hbar \omega/ [\exp(\hbar \omega / k_{\rm B}T_i) -1]$ describes the
energy of the thermally occupied photonic states with frequency $\omega$ and wave vector parallel to the interfaces in the multilayers 
${\bf k} = (k_x,k_y)$. The function $\tau(\omega,{\bf k},d)$ is the total transmission probability of the electromagnetic propagating 
waves ($|{\bf k}|=k < \omega/c$), as well as evanescent ones ($k> \omega/c$), and is expressed as \cite{Biehs2011,Moncada-Villa2015}
\begin{eqnarray}
\label{eq-trans-man}
\tau(\omega,{\bf k},d) = \hspace{7cm} & & \\ \left\{ \begin{array}{ll}
\mbox{Tr} \left\{ [\hat 1 - \hat {\cal R}_{21} \hat {\cal R}^{\dagger}_{21} ] \hat {\cal D}^{\dagger}
[\hat 1 - \hat {\cal R}^{\dagger}_{23} \hat {\cal R}_{23} ] \hat {\cal D} \right\}, & k < \omega/c \\
\mbox{Tr} \left\{ [\hat {\cal R}_{21} - \hat {\cal R}^{\dagger}_{21} ] \hat {\cal D}^{\dagger}
[\hat {\cal R}^{\dagger}_{23} - \hat {\cal R}_{23} ] \hat {\cal D} \right\} e^{-2|q_2|d}, & k > \omega/c
\end{array} \right. . & & \nonumber
\end{eqnarray}
Here, $q_2 = \sqrt{\omega^2/c^2 - k^2}$ is the $z$-component of the wave vector in the vacuum gap, $c$ is the velocity of light in 
vacuum and $\hat {\cal D}= [ \hat 1 - \hat {\cal R}_{21} \hat {\cal R}_{23} e^{2iq_2d} ]^{-1}$ describes the usual Fabry-P\'erot-like 
denominator resulting from the multiple scattering in the vacuum gap between the two multilayered systems. The $2 \times 2$ matrices 
$\hat {\cal R}_{ij}$ are the reflections matrices characterizing the two interfaces at both sides of the gap, and contain the 
information of the multiple scattering processes taking place inside each structure. These matrices have the following generic structure
\begin{equation}
\label{refl-mat}
\hat {\cal R}_{ij} = \left( \begin{array}{cc} r^{s,s}_{ij} & r^{s,p}_{ij} \\ 
r^{p,s}_{ij} & r^{p,p}_{ij} \end{array} \right) ,
\end{equation}
where $r^{\alpha, \beta}_{ij}$ with $\alpha,\beta =s,p$ (or TE, TM) is the reflection amplitude for the scattering of an incoming 
$\alpha$-polarized plane wave into an outgoing $\beta$-polarized wave. In practice, we compute numerically the different reflection 
matrices appearing in Eq.~(\ref{eq-trans-man}) by using the scattering-matrix approach for anisotropic multilayer systems that is 
described in Refs.~\cite{Caballero2012,Moncada-Villa2015}. It is worth remarking that the approach used in this work, contrary to the 
effective medium theory \cite{Liu2014,Song2020}, provides the exact numerical results for this problem and it can be applied for
an arbitrary number of unit cells in the 1D crystals. Let us also say that all the results presented in the work for the heat transfer 
coefficient were obtained for room temperature ($T=300$ K).

\section{Zero-field radiative heat transfer} \label{sec-zero}

\begin{figure}[t]
\includegraphics[width=\columnwidth,clip]{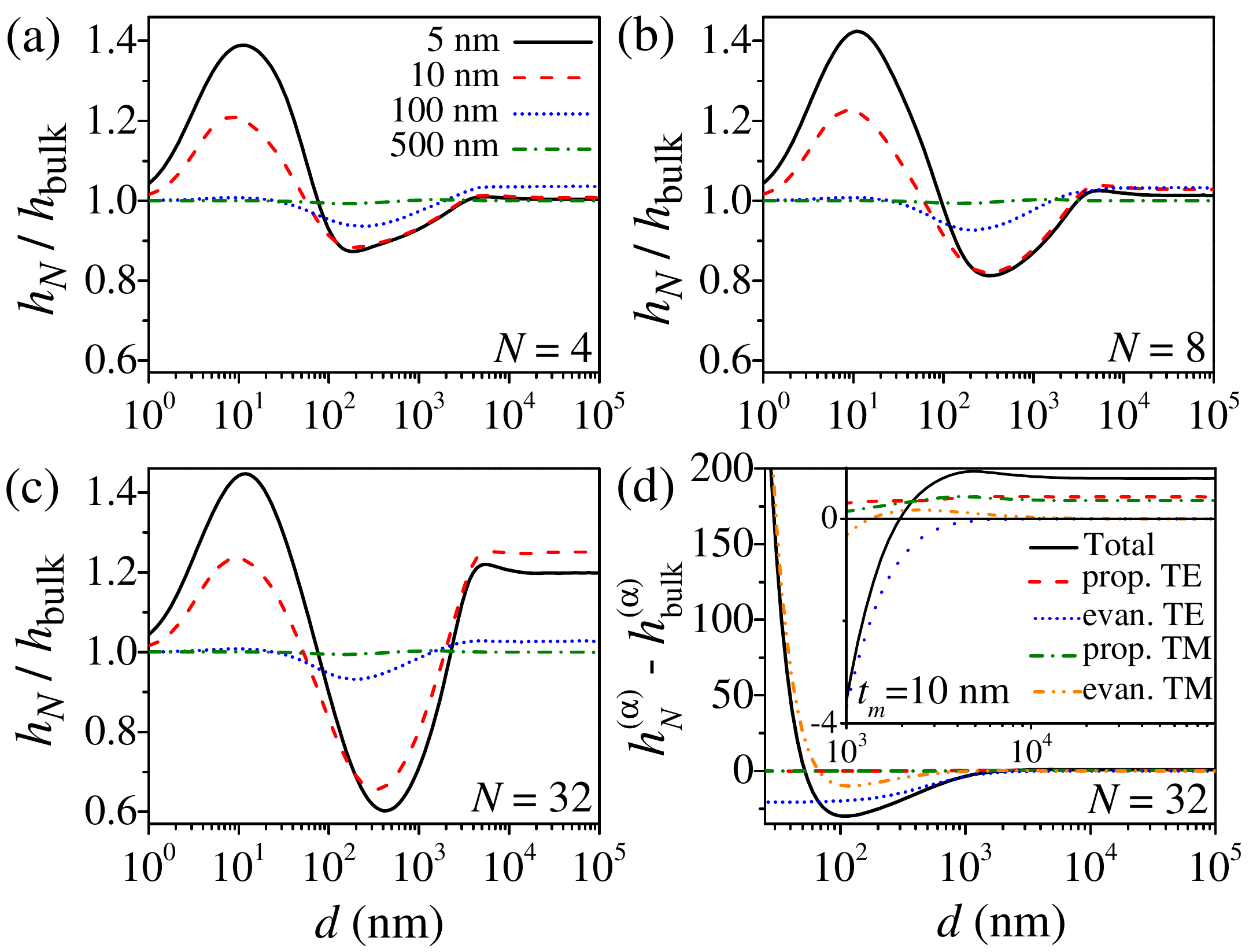}
\caption{(a-c) Ratio between the zero-field heat transfer coefficient of two multilayers, $h_N$, and the heat transfer coefficient
of two semi-infinite plates of InSb, $h_{\rm bulk}$, as a function of the gap size $d$. Panels (a), (b) and (c), correspond to 
$N=4$, $N=8$ and $N=32$ unit cells, respectively, and the different lines in each graph correspond to different values of 
thickness of the MO layers of InSb: 5 nm (solid line), 10 nm (dashed line), 100 nm (dotted line), and 500 nm (dash dotted line).
The dielectric material is BK7 and its layers have a thickness of 10 nm. (d) Gap dependence of the differences 
$h_N^{(\alpha)}-h_{\rm bulk}^{(\alpha)}$, where $\alpha$ stands for total, propagating and evanescent TE or TM, for $N=32$ unit cells, 
and $t_m = $10 nm [i.e., for the dashed line in panel (c)].}
\label{fig-zero-heat}
\end{figure}

We start the discussion of our results by first considering the radiative heat transfer between the multilayer systems of 
Fig.~\ref{fig-system}(a) in the absence of an external magnetic field. Radiative heat transfer in the near-field regime for this 
kind of systems, in absence of an external magnetic field, has already been reported in the literature \cite{Biehs2017,Iizuka2018}. 
In these works, it was shown that it is possible to enhance the NFRHT exchanged by two periodic multilayers, $h_N$, relative to the 
heat transfer between two infinite plates $h_{\rm bulk}$. Such an enhancement is a consequence of the hybridization of SPP and
SPhP modes with the Bloch modes resulting from the translation symmetry in a periodic structure. However, no much attention has been 
paid to the radiative heat transfer in the intermediate and far-field regimes. With this in mind, we start by presenting in 
Fig.~\ref{fig-zero-heat}(a-c) the heat transfer coefficient for the two multilayer structures, $h_N$, as a function of the gap size $d$, 
normalized to the heat transfer coefficient for two InSb infinite plates, $h_{\rm bulk}$, for three different values of the number 
of unit cells in each structure, $N$, and for several values of the thickness of the InSb layers, $t_m$. In this case we assumed that
the dielectric layers were made of BK7 glass. For gap sizes between 10 nm to 80 nm we observe an enhancement of the NFRHT, which becomes
more prominent as the thickness of the InSb layers is reduced and it reaches a maximum at $d \sim$ 10 nm. This is in qualitative agreement 
with what it has been reported in literature for layered structures with metal/vacuum unit cells \cite{Iizuka2018}. However, for 
gap sizes larger than about 80 nm, this enhancement disappears up to gaps on the order of 2200 nm and the radiative heat flux between 
the multilayer systems is actually smaller than in the case of two infinite parallel plates of InSb. Beyond gap sizes of about 2200 nm, 
an in particular in the far-field regime, the radiative heat flux between multilayered stacks is again larger than the corresponding 
one for two InSb parallel plates. 

\begin{figure*}[t]
\includegraphics[width=\textwidth,clip]{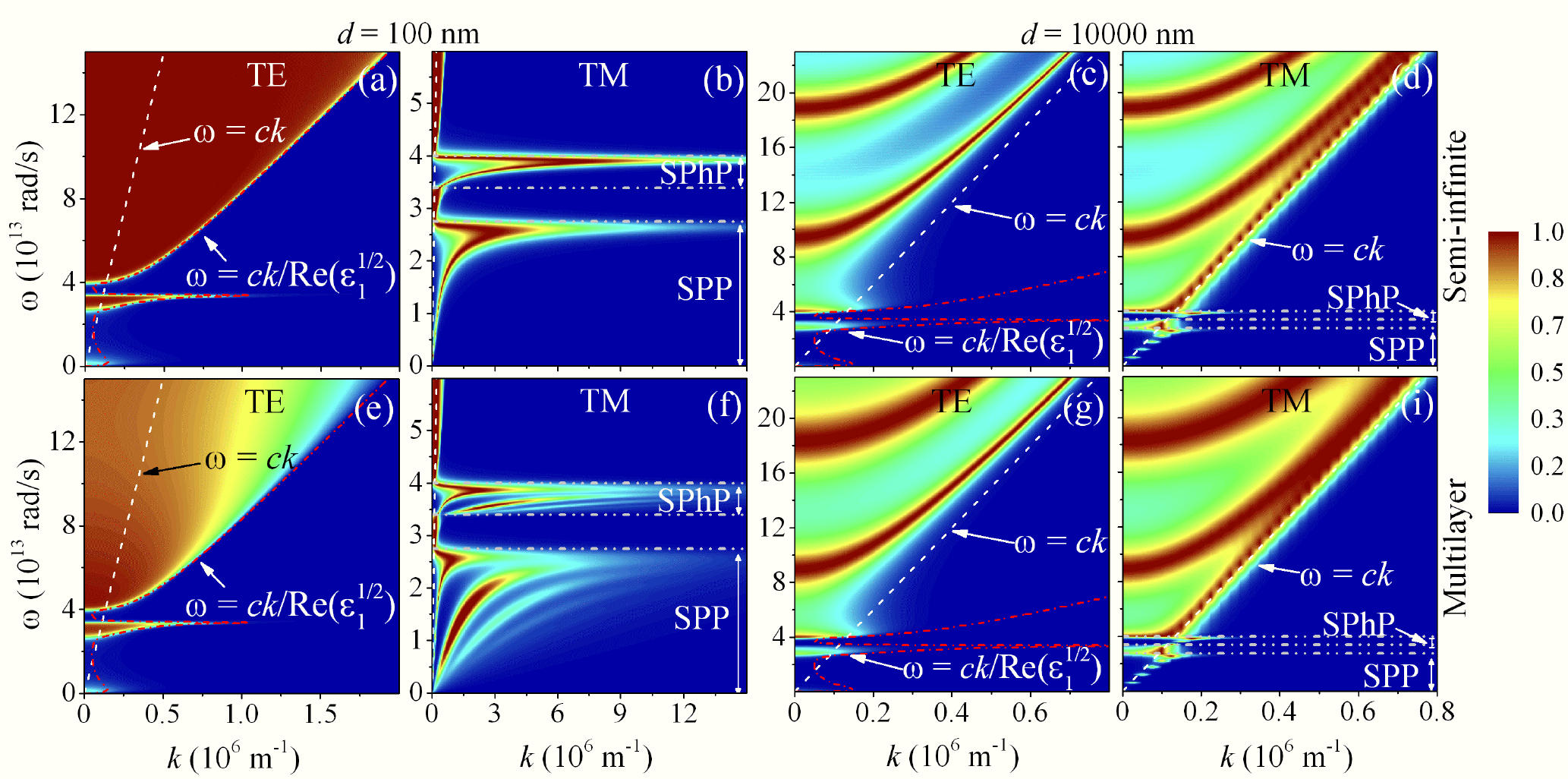}
\caption{Transmission coefficient for $s$-polarized (TE) [panels (a), (c), (e) and (g)] and $p$-polarized (TM) [panels (b), (d), (f) 
and (i)] waves, as a function of the magnitude of the parallel wave vector and frequency, in absence of an external magnetic field. 
Panels (a), (b), (e) and (f), correspond to a gap size $d=100$ nm, whereas the rest to a gap size of $d =10000$ nm. The white dashed 
lines indicate the light line in vacuum, $\omega=ck$, while the red dashed-dotted lines in panels (a), (c), (e) and (g) correspond 
to the light line inside InSb, $\omega =ck /\mbox{Re}(\epsilon^{1/2}_{1})$. The horizontal dashed dotted lines in panels (b), (d), (f) and (i), 
delimit the different regions where SPP and SPhP modes exist for the TM polarization. The upper panels correspond to the case of two 
InSb semi-infinite plates, whereas the lower ones are for a multilayer system comprising of $N=32$ unit cells of InSb/BK7, with 
$t_m=t_d=$ 10 nm, as in Fig.~\ref{fig-zero-heat}(d).}
\label{fig-tran}
\end{figure*}
\begin{figure}[t]
\includegraphics[width=\columnwidth,clip]{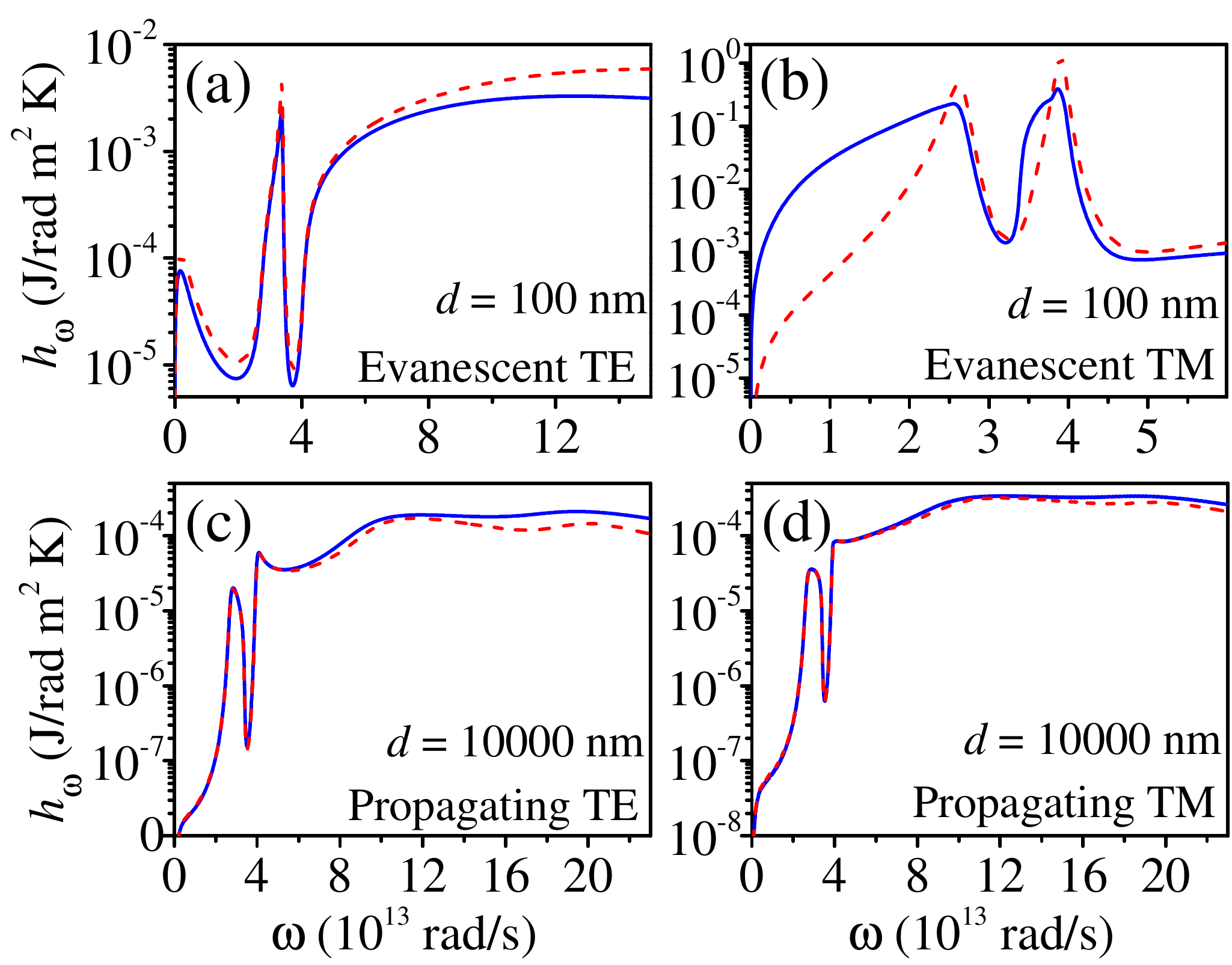}
\caption{Comparison between spectral heat transfer coefficients of two InSb infinite plates (dashed lines) and the 
corresponding one for the two multilayers of Fig.~\ref{fig-system}(a) (solid lines), with $N=32$ InSb/BK7 unit cells, 
and $t_m=t_d=$ 10 nm, as in Fig.~\ref{fig-zero-heat}(d). Panels (a), (b), (c) and (d) correspond, respectively, to the pair 
of panels [(a) and (e)], [(b) and (f)], [(c) and (g)] and [(d) and (i)], in Fig.~\ref{fig-tran}.}
\label{fig-spec-zero}
\end{figure}

In order to understand these new features for gap sizes above 80 nm, we present in Fig.~\ref{fig-zero-heat}(d) the difference between 
the total heat transfer coefficients $h_N$ and $h_{\rm bulk}$, and between each one of their separate contributions (evanescent and 
propagating waves for TE and TM polarizations). This is done for the case represented by a dashed line in panel (c). As it can be seen, 
the enhancement in the near-field regime ($d \leq 80$ nm) is entirely due to the contribution of the TM evanescent modes associated to 
the SPP and SPhPs modes. We also see that the decrease of the heat flux $h_N$ relative to $h_{\rm bulk}$ for gap sizes between 80 nm 
and 2200 nm is due to the decrease of both TE and TM evanescent contributions to the total flux. To provide some physical insight for 
the decrease of these contributions, we present in Fig.~\ref{fig-tran} the comparison between the transmission probabilities corresponding 
to the multilayer configuration and the corresponding ones for two infinite InSb plates, for TE and TM polarizations and for two different 
values of the gap size. We observe from panels (a) and (e) that the electromagnetic modes on the right of the vacuum light line $\omega=ck$ 
and on the left of the InSb light line $\omega =ck/ \mbox{Re}(\epsilon^{1/2}_{1})$, i.e., the frustrated TE evanescent modes, are more 
probable for the case of two infinite plates than for the multilayer system. This change is due to the fact that the outermost layer in 
the stratified media has less propagating modes involved in total internal reflection at the interface with the vacuum gap. This decrease 
in the contribution of frustrated TE modes is indeed evident in Fig.~\ref{fig-spec-zero}(a), where we have plotted the spectral heat 
transfer contribution, $h_\omega$, corresponding to the transmission probabilities of panels (a) and (e) in Fig.~\ref{fig-tran}. Let us 
clarify that this spectral heat transfer coefficient is defined as the heat transfer coefficient per unit of frequency. Concerning the 
TM evanescent waves, we can see from panels (b) and (f) of Fig.~\ref{fig-tran} that transmission for the multilayer structures exhibits 
a higher number of surface states as a consequence of the hybridization of SPP and SPhP modes with Bloch modes. However, these hybridized 
modes have smaller values of the parallel component of the wave vector $k$ than in the case of two infinite plates. The reason for this 
is that as the gap size increases, the evanescent field of deepest layers in one structure cannot reach the vacuum gap and penetrate into 
the first layer of opposite structure. Then, as it can be observed in Fig.~\ref{fig-spec-zero}(b), the corresponding spectral heat transfer
coefficient has broader, but smaller central peaks than in the configuration of two infinite plates.  

We end this section by briefly discussing the increase of the far-field radiative heat transfer, see inset in Fig.~\ref{fig-zero-heat}. 
It is clear that such an increase originates from both TE and TM propagating modes, which in turn is a consequence of the enhanced intensity 
of Fabry-P\'erot resonances inside all the $2N$ layers of each multilayer system, as can be seen upon comparing panels (c) and (g) 
[(d) and (i)] for TE (TM) polarized waves, and in their corresponding spectral heat transfer coefficients in Fig.~\ref{fig-spec-zero}(c) 
[\ref{fig-spec-zero}(d)]. 

\section{Magnetic field effect on the radiative heat transfer} \label{sec-finite}

\begin{figure}[t]
\includegraphics[width=\columnwidth,clip]{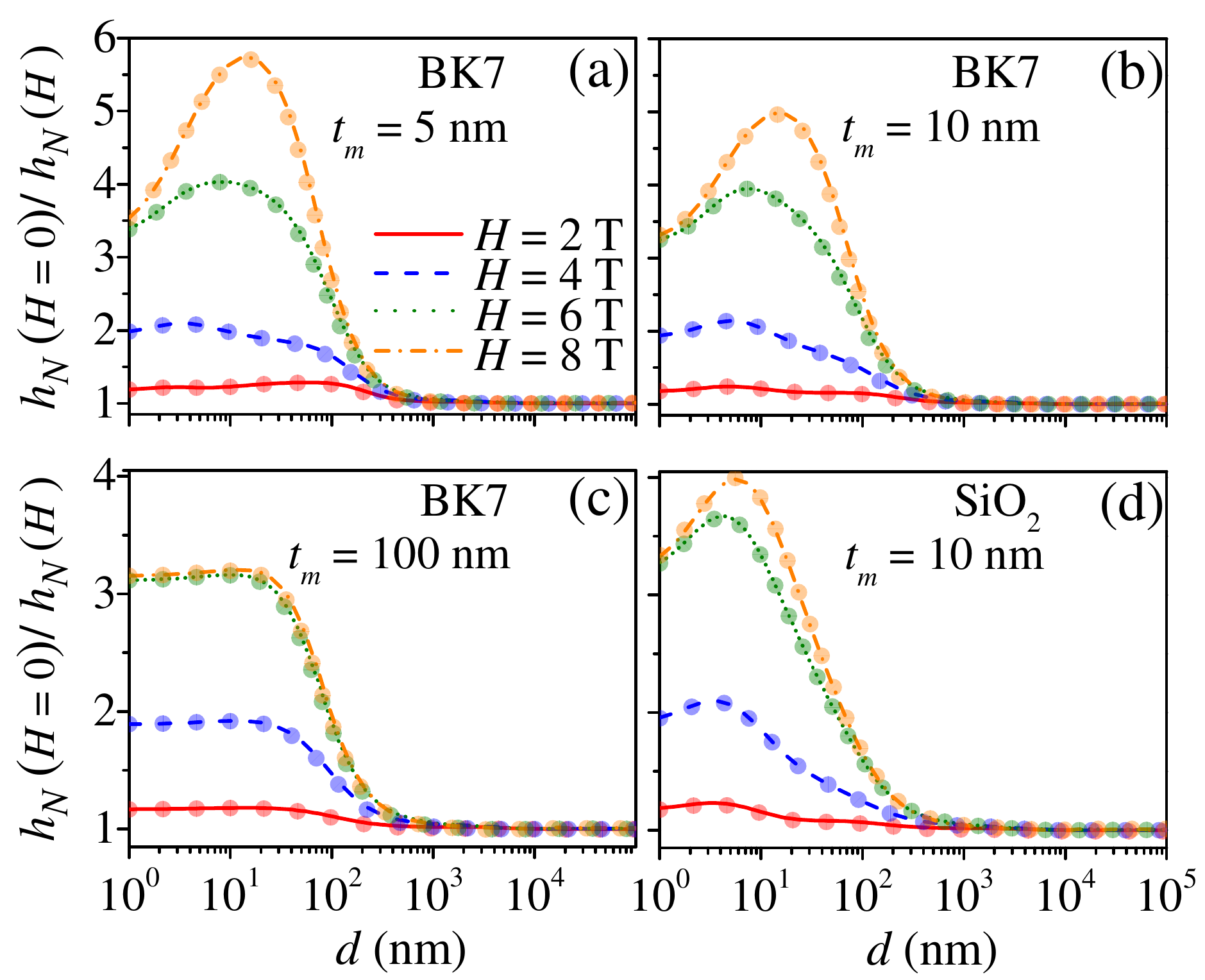}
\caption{Ratio between the zero-field heat transfer coefficient and the coefficient at a given value of the field magnitude 
as a function of the gap size and thickness, $t_m$, of the InSb layers. Panels (a)-(c) correspond to an InSb/BK7 unit cell, whereas 
panel (d) corresponds to an InSb/SiO$_2$ unit cell. All results were obtained for 16 unit cells and for dielectric layers (either BK7 
or SiO$_2$) with thickness $t_d=$ 10 nm. Circles in panels (a)-(c) correspond to the results obtained with the diagonal approximation 
$r_{ij}^{s,p}=r_{ij}^{p,s}=0$.}
\label{fig-heat}
\end{figure}

We turn now to discuss how the magnetic field affects the radiative heat transfer in the system under study. In Fig.~\ref{fig-heat}
we present the gap dependence of the ratio between the zero-field heat transfer coefficient and the corresponding coefficient at a given 
magnitude of the magnetic field for three different values of InSb layer thickness $t_m$ and $N=16$. Panels (a-c) display results for the
case in which the dielectric layer is made of BK7 glass, while panel (d) shows an example in which this layer is made of SiO$_2$. The
first thing to notice is that the applied magnetic field always reduces the heat flux (i.e., $h_N(H=0)/h_N(H)>1$), irrespective
of the dielectric material or the values of the geometrical parameter. Notice also that this field-induced reduction is much more
pronounced in the near-field regime. As explained in the introduction, this heat transfer reduction is indeed the 
expected result from our previous studies of the case of two InSb parallel plates \cite{Moncada-Villa2015} and a plate of InSb 
separated by a vacuum gap from a InSb thin film \cite{Moncada-Villa2020}. These results are clearly at variance with those reported 
in Ref.~\cite{Song2020} that were obtained with the help of an effective medium theory. It is worth stressing that we have found very 
similar results for different numbers of unit cells (not shown here). 

\begin{figure}[h!]
\includegraphics[width=\columnwidth,clip]{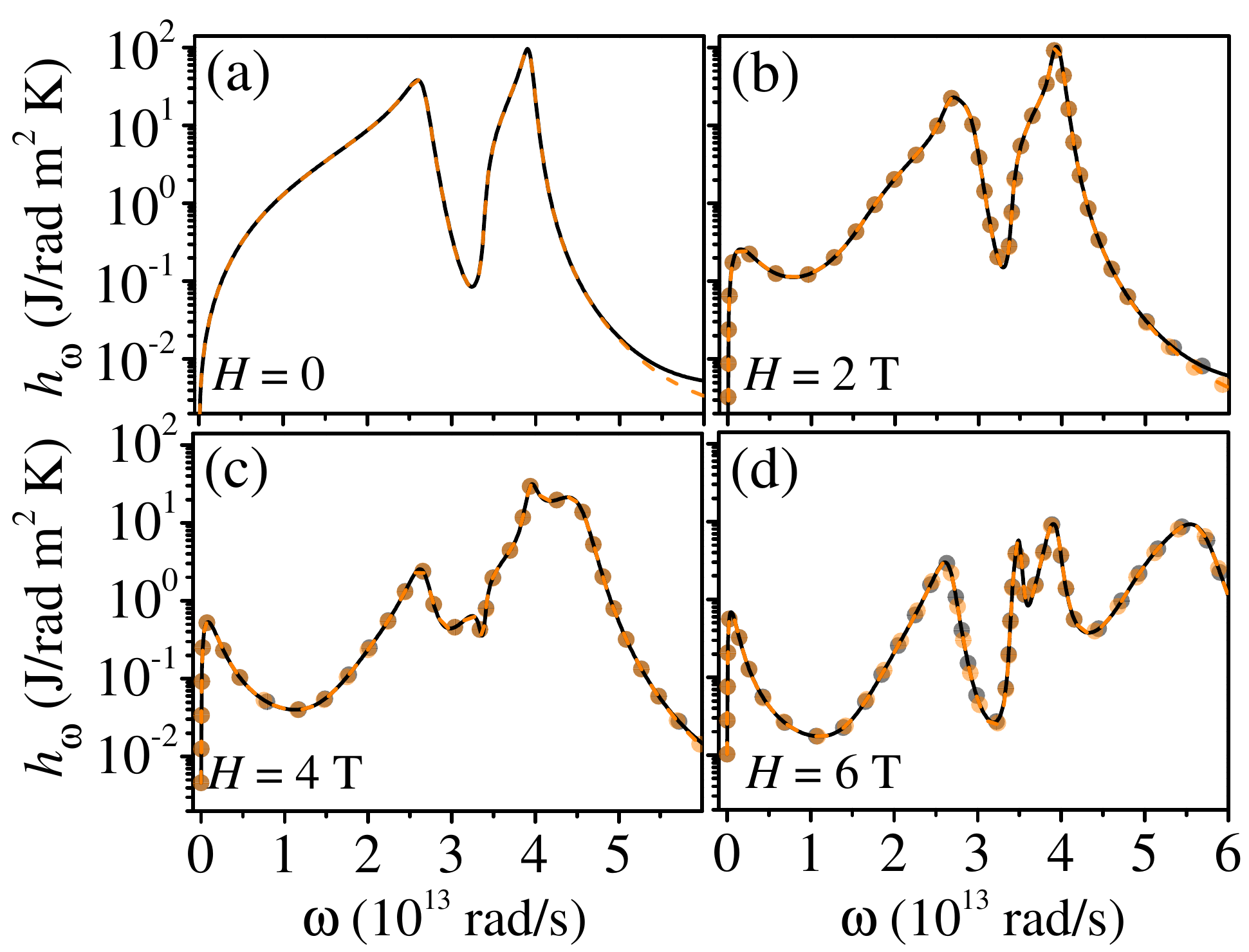}
\caption{Total spectral heat transfer coefficient (black solid lines) and the contribution of TM evanescent waves (dashed orange lines)
for the same geometrical parameters as in Fig.~\ref{fig-heat}(b) and a gap size of 10 nm. The different panels correspond to
different values of the external magnetic field: (a) 0 T, (b) 2 T, (c) 4 T, and (d) 6 T. The number of unit cells in all cases shown in
this figure were set to $N=16$. The circles in panels (b)-(d) correspond to the results for the total spectral heat transfer coefficient 
obtained with the diagonal approximation $r_{ij}^{s,p}=r_{ij}^{p,s}=0$.}
\label{fig-spec-fin}
\end{figure}
\begin{figure*}[t]
\includegraphics[width=\textwidth,clip]{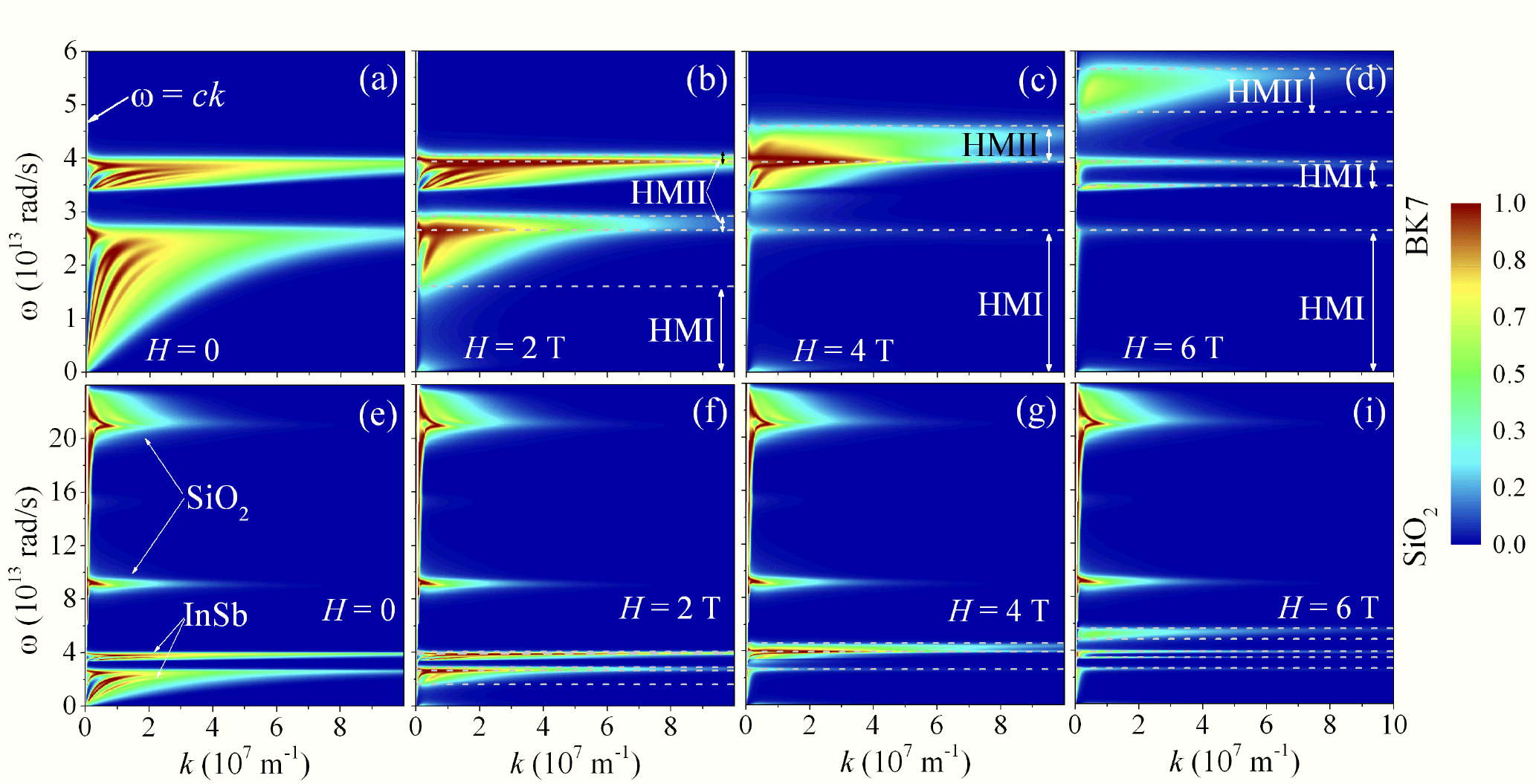}
\caption{Transmission coefficient for $p$-polarized (TM) waves as a function of the magnitude of the parallel wave vector and 
frequency for different magnitudes of the applied magnetic field. Upper (lower) panels correspond to an InSb/BK7 (InSb/SiO$_2$) unit 
cell. In all panels we have set $t_m=t_d=d=10$ nm. The number of units cells was set to $N=16$.}
\label{fig-trans-fin}
\end{figure*}

The results presented in Fig.~\ref{fig-heat} show that the most drastic NFRHT reduction induced by the magnetic field occurs for the 
thinnest InSb layer. For example, we see in panel (a) that for a 5 nm-thick InSb layer the NFRHT is reduced by up to a 573\% for a gap 
size of 14.4 nm and a field magnitude of 8 T. As the InSb layer gets thicker, for example 10 nm as in panel (b), the maximum reduction
becomes on the order of $\sim 500$\%, and for a thickness of the InSb layer of 100 nm, see panel (c),  one essentially recovers the limiting 
case of two infinite plates discussed in Ref.~\cite{Moncada-Villa2015}. The reason for this behavior is easy to understand in view of 
our previous studies \cite{Moncada-Villa2015,Moncada-Villa2020}. For thin InSb layers, the number of these layers whose evanescent field 
reaches the vacuum gap is larger and, therefore, the replacement of SPPs and SPhPs for modes by hyperbolic ones has a more drastic impact.  

A natural question that arises at this point is if the polarization conversion plays any role in the decease of the NFRHT
when the magnetic field is applied. To answer this question we have computed the heat transfer in all the examples discussed above,
but this time assuming that the the amplitudes $r_{ij}^{p,s}$ and $r_{ij}^{s,p}$ are zero. Let us recall that these off-diagonal 
reflection coefficients, see Eq.~\eqref{refl-mat}, are responsible for the polarization conversion in this system. The results 
obtained with this approximation are shown in Fig.~\ref{fig-heat} as circles. As it is evident from these results, the polarization
conversion plays no role in the reduction of the NFRHT induced by the external magnetic field. As discussed in our previous work
\cite{Moncada-Villa2015,Moncada-Villa2020}, this reduction is due to the fact that the field induces the appearance of hyperbolic
modes that progressively replace the zero-field surface modes and they are less effective transferring the heat. A way to confirm this 
argument is by investigating the spectral heat transfer coefficient, which we show in Fig.~\ref{fig-spec-fin} for the case of a gap 
size of 10 nm shown in Fig.~\ref{fig-heat}(b) and values of the field equal to 0, 2, 4, and 6 T. In this figure we also show the contribution 
of the TM evanescent waves and the results for the total heat transfer coefficient computed with the diagonal approximation 
$r_{ij}^{p,s}=r_{ij}^{s,p}=0$. The first thing to notice from these spectral functions is that, as we showed in our previous papers 
\cite{Moncada-Villa2015,Moncada-Villa2020}, the reduction of the NFRHT is dominated exclusively by the TM polarized evanescent modes, 
which are suppressed by the emerging hyperbolic modes as the field is cranked up. The second thing is that the results obtained with the
diagonal approximation reproduce very accurately the exact results, showing again that the polarization conversion is not relevant for
the field-induced NFRHT reduction.

Another interesting observation is the fact that the NFRHT reduction induced by the field is more prominent in the case in which the
dielectric layer is made of BK7 glass, which is evident when comparing panels (b) and (d) of Fig.~\ref{fig-heat} that were obtained
for the same value of the dielectric layer thickness. In particular, we find that for an applied field with 8 T and a gap size of 10 nm, 
the reduction for the BK7 case is about 488\%, while it is 381\% for SiO$_2$. This difference can be understood by examining the 
evolution of the transmission probability of the TM evanescent modes as the magnitude of the external field varies, see 
Fig.~\ref{fig-trans-fin}. In the structure with InSb/BK7 unit cells, the absence of dispersion in the dielectric function of the BK7 layers
makes that SPPs and SPhPs modes of InSb be the only ones available to hybridize with the Bloch modes [see panel (a)]. These modes are 
progressively replaced by hyperbolic modes as the external field increases \cite{Moncada-Villa2015}, leading to the reduction of the 
NFRHT [see panels (b)-(d)]. In contrast, when the dielectric layers are made of a polar material like SiO$_2$, the zero-field transmission
probability exhibits, in addition to the maxima related to the SPPs and SPhPs of the InSb layers, other two maxima at higher frequencies
related the SPhPs in the SiO$_2$ layers [see Fig.~\ref{fig-system}(d)]. So, although the external field replaces the SPP and SPhP modes 
of InSb by hyperbolic modes, this field cannot modify the SPhP modes due to SiO$_2$ [see panels (f) to (i)], making the structure containing
this polar dielectric less sensitive to the application of an external magnetic field.

\section{Conclusions} \label{sec-conclusions}

Motivated by the current interest in the active control of NFRHT, we have presented in this paper a systematic 
theoretical study of the influence of an external magnetic field in the radiative heat exchange between two periodic 1D 
magneto-photonic crystals. We have analyzed in detail both the zero- and the finite-field radiative heat flux. In the first case, 
we have shown that for the smallest gaps in the near-field regime, the heat flux between the multilayered structures is higher
than for the case of two InSb infinite parallel plates. In agreement with related studies, we find that this enhancement is due
to the hybridization of SPP and SPhP modes with the Bloch modes resulting from the translation symmetry in a periodic structure.
However, in an intermediate regime for gap sizes between 80 and 2200 nm, the radiative heat transfer is actually smaller in the case of the 
1D crystals. We have shown that this is a consequence of a decrease in the efficiency of frustrated TE modes by total internal reflection, 
and smaller parallel wave vectors for surface TM waves. In contrast, for gap sizes beyond 2200 nm and, in particular in the far-field 
regime, there is an increase in the radiative heat flux due to enhanced Fabry-P\'erot resonances between and inside multiple layers of 
each stack. 

On the other hand, we have shown that an applied static magnetic field induces the replacement of surface polariton states, which 
have very high values of the parallel wave vector, by hyperbolic modes with lower parallel wave vectors. The field-induced appearance
of these hyperbolic modes leads to a reduction of the NFRHT that is even more pronounced than in the case of two InSb infinite parallel plates. 
In particular, we have found reduction factors as high as 6 for fields of a few Teslas, which is truly remarkable. Moreover, we have shown 
that this reduction of the NFRHT is due to the appearance of hyperbolic modes induced by the field and that the polarization conversion 
in this system plays no role in this effect. Overall, our work provides one more example of the promising possibility to control NFRHT by 
means of an external magnetic field, and we hope that it will motivate the realization of experiments to confirm these interesting 
thermomagnetic effects.

\acknowledgments

J.C.C.\ acknowledges funding from the Spanish Ministry of Economy and Competitiveness (MINECO) (contract No.\ FIS2017-84057-P).


\begin{thebibliography}{00}

\bibitem{Song2015a}
B. Song, A. Fiorino, E. Meyhofer, and P. Reddy,
Near-field radiative thermal transport: From theory to experiment,
AIP Advances {\bf 5}, 053503 (2015).

\bibitem{Cuevas2018}
J.~C. Cuevas and F.~J. Garc\'{\i}a-Vidal, 
Radiative heat transfer, ACS Photonics {\bf 5}, 3896 (2018).

\bibitem{Biehs2020}
S.-A. Biehs, R. Messina, P.~S. Venkataram, A.~W. Rodriguez, J.~C. Cuevas, P. Ben-Abdallah,
Near-field radiative heat transfer in many-body systems, arXiv:2007.05604.

\bibitem{Polder1971}
D. Polder and M. Van Hove, 
Theory of radiative heat transfer between closely spaced bodies
Phys. Rev. B {\bf 4}, 3303 (1971).

\bibitem{Kittel2005}
A. Kittel, W.~M\"uller-Hirsch, J. Parisi, S.-A. Biehs, D. Reddig, and M. Holthaus,
Near-Field Heat Transfer in a Scanning Thermal Microscope,
Phys. Rev. Lett. {\bf 95}, 224301 (2005).

\bibitem{Narayanaswamy2008}
A. Narayanaswamy, S. Shen, and G. Chen,
Near-field radiative heat transfer between a sphere and a substrate,
Phys. Rev. B {\bf 78}, 115303 (2008).

\bibitem{Hu2008}
L. Hu, A. Narayanaswamy, X.~Y. Chen, and G. Chen, 
Near-field thermal radiation between two closely spaced glass plates
exceeding Planck’s blackbody radiation law,
Appl. Phys. Lett. {\bf 92}, 133106 (2008).

\bibitem{Rousseau2009}
E. Rousseau, A. Siria, G. Jourdan, S. Volz, F. Comin, J. Chevrier, and J.-J, Greffet,
Radiative heat transfer at the nanoscale,
Nat. Photonics {\bf 3}, 514 (2009).

\bibitem{Shen2009}
S. Shen, A. Narayanaswamy, and G. Chen, 
Surface phonon polaritons mediated energy transfer between nanoscale gaps,
Nano Lett. {\bf 9}, 2909 (2009).

\bibitem{Ottens2011}
R.~S. Ottens, V. Quetschke, S. Wise, A.~A. Alemi, R. Lundock, G. Mueller, D.~H. Reitze, 
D.~B. Tanner, and B.~F. Whiting,
Near- Field Radiative Heat Transfer Between Macroscopic Planar Surfaces,
Phys. Rev. Lett. {\bf 107}, 014301 (2011).

\bibitem{Shen2012}
S. Shen, A. Mavrokefalos, P. Sambegoro, and G. Chen,
Nanoscale thermal radiation between two gold surfaces,
Appl. Phys. Lett. {\bf 100}, 233114 (2012).

\bibitem{Kralik2012}
T. Kralik, P. Hanzelka, M. Zobac, V. Musilova, T. Fort, and M. Horak, 
Strong Near-Field Enhancement of Radiative Heat Transfer Between Metallic Surfaces,
Phys. Rev. Lett. {\bf 109}, 224302 (2012).

\bibitem{Zwol2012a}
P.~J. van Zwol, L. Ranno, and J. Chevrier,
Tuning Near Field Radiative Heat Flux Through Surface Excitations with a Metal Insulator Transition,
Phys. Rev. Lett. {\bf 108}, 234301 (2012).

\bibitem{Zwol2012b}
P.~J. van Zwol, S. Thiele, C. Berger, W.~A. de Heer, and J. Chevrier, 
Nanoscale Radiative Heat Flow Due to Surface Plasmons in Graphene and Doped Silicon,
Phys. Rev. Lett. {\bf 109}, 264301 (2012).

\bibitem{Guha2012}
B. Guha, C. Otey, C.~B. Poitras, S.~H. Fan, and M. Lipson, 
Near-field radiative cooling of nanostructures,
Nano Lett. {\bf 12}, 4546 (2012).

\bibitem{Shi2013}
J. Shi, P. Li, B. Liu, and S. Shen, 
Tuning near field radiation by doped silicon,
Appl. Phys. Lett. {\bf 102}, 183114 (2013).

\bibitem{Worbes2013}
L. Worbes, D. Hellmann, and A. Kittel, 
Enhanced near-field heat flow of a monolayer dielectric island,
Phys. Rev. Lett. {\bf 110}, 134302 (2013).

\bibitem{St-Gelais2014}
R. St-Gelais, B. Guha, L.~X. Zhu, S.~H. Fan, and M. Lipson, 
Demonstration of strong near-field radiative heat transfer between
integrated nanostructures,
Nano Lett. {\bf 14}, 6971 (2014).

\bibitem{Song2015b}
B. Song, Y. Ganjeh, S. Sadat, D. Thompson, A. Fiorino, V. Fern\'andez-Hurtado, 
J. Feist, F.~J. Garcia-Vidal, J.~C. Cuevas, P. Reddy, and E. Meyhofer, 
Enhancement of near-field radiative heat transfer using polar dielectric thin films,
Nat. Nanotechnol. {\bf 10}, 253 (2015).

\bibitem{Kim2015}
K. Kim, B. Song, V. Fern\'andez-Hurtado, W. Lee, W. Jeong, L. Cui, D. Thompson, 
J. Feist, M.~T.~H. Reid, F.~J. Garc\'{\i}a-Vidal, J.~C. Cuevas, E. Meyhofer, P. Reddy,
Radiative heat transfer in the extreme near field,
Nature (London) {\bf 528}, 387 (2015).

\bibitem{Lim2015}
M. Lim, S.~S. Lee, and B.~J. Lee,
Near-field thermal radiation between doped silicon plates at nanoscale gaps,
Phys. Rev. B {\bf 91}, 195136 (2015).

\bibitem{St-Gelais2016}
R. St-Gelais, L. Zhu, S. Fan, and M. Lipson,
Near-field radiative heat transfer between parallel structures in the deep 
subwavelength regime,
Nat. Nanotechnol. {\bf 11}, 515 (2016).

\bibitem{Song2016}
B. Song, D. Thompson, A. Fiorino, Y. Ganjeh, P. Reddy, E. Meyhofer,
Radiative heat conductances between dielectric and metallic parallel 
plates with nanoscale gaps,
Nat. Nanotechnol. {\bf 11}, 509 (2016).

\bibitem{Bernardi2016}
M.~P. Bernardi, D. Milovich, M. Francoeur,
Radiative heat transfer exceeding the blackbody limit between 
macroscale planar surfaces separated by a nanosize vacuum gap,
Nat. Commun. {\bf 7}, 12900 (2016).

\bibitem{Cui2017}
L. Cui, W. Jeong, V. Fern\'andez-Hurtado, J. Feist, F.~J. Garc\'{\i}a-Vidal, 
J.~C. Cuevas, E. Meyhofer, P. Reddy,
Study of radiative heat transfer in {\AA}ngstr\"om- and nanometre-sized gaps,
Nat. Commun. {\bf 8}, 14479 (2017).

\bibitem{Kloppstech2017} 
K. Kloppstech, N. K\"onne, S.-A. Biehs, A.~W. Rodriguez, L. Worbes, D. Hellmann, A. Kittel, 
Giant heat transfer in the crossover regime between conduction and radiation, 
Nat. Commun. {\bf 8}, 14475 (2018).

\bibitem{Ghashami2018}
M. Ghashami, H. Geng, T. Kim, N. Iacopino, S.-K. Cho, K. Park,
Precision Measurement of Phonon-Polaritonic Near-Field Energy Transfer Between Macroscale 
Planar Structures Under Large Thermal Gradients,
Phys. Rev. Lett. {\bf 120}, 175901 (2018).

\bibitem{Fiorino2018}
A. Fiorino, D. Thompson, L. Zhu, B. Song, P. Reddy, E. Meyhofer,
Giant enhancement in radiative heat transfer in sub-30 nm gaps of plane parallel surfaces,
Nano Lett. {\bf 18}, 3711 (2018).

\bibitem{DeSutter2019}
J. DeSutter, L. Tang, and M. Francoeur,
A near-field radiative heat transfer device,
Nat. Nanotechnol. {\bf 14}, 751 (2019).

\bibitem{Rytov1953}
S.~M. Rytov, \emph{Theory of Electric Fluctuations and Thermal Radiation},
(Air Force Cambrige Research Center, Bedford, MA, 1953).

\bibitem{Rytov1989}
S.~M. Rytov, Y.~A. Kravtsov, and V.~I. Tatarskii, \emph{Principles of Statistical Radiophysics},
Vol.\ 3 (Springer-Verlag, Berlin Heidelberg, 1989).

\bibitem{Moncada-Villa2015}
E. Moncada-Villa, V. Fern\'andez-Hurtado, F.~J. Garc\'{\i}a-Vidal, 
A. Garc\'{\i}a-Mart\'{\i}n, and J.~C. Cuevas,
Magnetic field control of near-field radiative heat transfer and the realization 
of highly tunable hyperbolic thermal emitters,
Phys. Rev B {\bf 92}, 125418 (2015).

\bibitem{Ben-Abdallah2016a}
P. Ben-Abdallah,
Photon Thermal Hall Effect,
Phys. Rev. Lett. {\bf 116}, 084301 (2016).

\bibitem{Zhu2016}
L. Zhu and S. Fan, 
Persistent Directional Current at Equilibrium in Nonreciprocal Many-Body
 Near Field Electromagnetic Heat Transfer,
 Phys. Rev. Lett. {\bf 117}, 134303 (2016).

\bibitem{Latella2017}
I. Latella and P. Ben-Abdallah,
Giant Thermal Magnetoresis- tance in Plasmonic Structures,
Phys. Rev. Lett. {\bf 118}, 173902 (2017).

\bibitem{Abraham-Ekeroth2018}
R.~M. Abraham Ekeroth, P. Ben-Abdallah, J.~C. Cuevas, and Garc\'{\i}a-Mart\'{\i}n,
Anisotropic thermal magnetoresistance for an active control of radiative heat transfer,
ACS Photonics {\bf 5}, 705 (2018).

\bibitem{Moncada-Villa2020}
E. Moncada-Villa and J. C. Cuevas, 
Magnetic field effects in the near-field radiative heat transfer between planar structures,
Phys. Rev. B {\bf 101}, 085411 (2020). 

\bibitem{Ott2019}
A. Ott, R. Messina, P. Ben-Abdallah, and S.-A. Biehs,
Magnetothermoplasmonics: from theory to applications, 
J. Photon. Energy {\bf 9}, 032711 (2019).

\bibitem{Moncada-Villa2019}
E. Moncada-Villa, A.~I. Fern\'andez-Dom\'{\i}nguez, J.~C. Cuevas,
Magnetic-field controlled anomalous refraction in doped semiconductors,
J. Opt. Soc. Am. B {\bf 36}, 935 (2019).

\bibitem{Song2020}
J. Song, Q. Cheng , L. Lu, B. Li, K. Zhou, B. Zhang, Z. Luo, and X. Zhou,
Magnetically tunable near-field radiative heat transfer in hyperbolic metamaterials,
Phys. Rev. Appl. {\bf 13} 024054 (2020). 

\bibitem{Guo2012}
Y. Guo, C.~L. Cortes, S. Molesky, and Z. Jacob, 
Broadband super-Planckian thermal emission from hyperbolic metamaterials,
Appl. Phys. Lett. {\bf 101}, 131106 (2012).

\bibitem{Biehs2012}
S.~A. Biehs, M. Tschikin, and P. Ben-Abdallah, 
Hyperbolic Metamaterials as an Analog of a Blackbody in the Near Field,
Phys. Rev. Lett. {\bf 109}, 104301 (2012).

\bibitem{Guo2013}
Y. Guo and Z. Jacob, 
Thermal hyperbolic metamaterials,
Opt. Express {\bf 21}, 15014 (2013).

\bibitem{Biehs2013}
S.-A. Biehs, M. Tschikin, R. Messina, and P. Ben-Abdallah,
Super-Planckian near-field thermal emission with phonon-polaritonic hyperbolic metamaterials,
Appl. Phys. Lett. {\bf 102}, 131106 (2013).

\bibitem{Bright2014}
T.~J. Bright, X.~L. Liu, and Z.~M. Zhang, 
Energy streamlines in near-field radiative heat transfer between hyperbolic metamaterials,
Opt. Express {\bf 22}, A1112 (2014).

\bibitem{Miller2014}
O.~D. Miller, S.~G. Johnson, and A.~W. Rodriguez, 
Effectiveness of Thin Films in Lieu of Hyperbolic Metamaterials in the Near Field,
Phys. Rev. Lett. {\bf 112}, 157402 (2014).

\bibitem{Biehs2017}
S.-A. Biehs and P. Ben-Abdallah, 
Near-field heat transfer between multilayer hyperbolic metamaterials,
Z. Naturforsch. A {\bf 72}, 115 (2017).

\bibitem{Iizuka2018}
H. Iizuka and S. Fan,
Significant enhancement of near-field electromagnetic heat transfer in a multilayer structure through 
multiple surface-states coupling,
Phys. Rev. Lett. {\bf 120}, 063901 (2018).

\bibitem{Fernandez-Hurtado2017}
V. Fern\'andez-Hurtado, F.~J. Garcia-Vidal, S. Fan, J.~C. Cuevas,
Enhancing Near-Field Radiative Heat Transfer with Si-Based Metasurfaces,
Phys. Rev. Lett. {\bf 118}, 203901 (2017).

\bibitem{Zvezdin1997}
A. Zvezdin and V. Kotov, 
\emph{Modern Magnetooptics and Magnetooptical Materials},
(IOP, Bristol, 1997).

\bibitem{Palik1976}
E.~D. Palik, R. Kaplan, R.~W. Gammon, H. Kaplan, R.~F. Wallis, and J.~J. Quinn, 
Coupled surface magnetoplasmon-optic-phonon polariton modes on InSb,
Phys. Rev. B {\bf 13}, 2497 (1976).

\bibitem{Palik1985}
E.~D. Palik, \emph{Handbook of Optical Constants of Solids}
(Academic Press, London, 1985).

\bibitem{Mulet2002}
J.~P. Mulet, K. Joulain, R. Carminati, and J.~J. Greffet,
Enhanced radiative heat transfer at nanometric distances, 
Microscale Therm. Eng. {\bf 6}, 209 (2002).

\bibitem{Biehs2011}
S.-A. Biehs, P. Ben-Abdallah, F.~S.S. Rosa, K. Joulain, and J.-J. Greffet, 
Nanoscale heat flux between nanoporous materials,
Opt. Express {\bf 19}, A1088 (2011).

\bibitem{Caballero2012}
B. Caballero, A. Garc\'{\i}a-Mart\'{\i}n, and J.~C. Cuevas,
Generalized scattering-matrix approach for magneto-optics in periodically patterned 
multilayer systems,
Phys. Rev. B {\bf 85}, 245103 (2012).

\bibitem{Liu2014}
X. L. Liu, T. J. Bright, and Z. M. Zhang,
Application conditions of effective medium theory in the near-field radiative heat transfer between 
multilayered metamaterials,
J. Heat Transfer {\bf 136}, 092703 (2014).

\end{thebibliography}
\end{document}